\documentclass[11pt]{article}
\usepackage[T2A]{fontenc}
\usepackage[utf8]{inputenc}
\usepackage[english]{babel}
\usepackage{graphicx}
\usepackage{wrapfig}
\usepackage{bmpsize}
\usepackage{amsmath,amsfonts,amssymb,amsthm,mathtools}
\usepackage{xcolor}
\usepackage{subcaption}
\usepackage{color}

\definecolor{BlueGreen}{RGB}{49,152,255}
\definecolor{Violet}{RGB}{120,80,120}
\definecolor{Blue}{RGB}{0,0,255}
\definecolor{Yellow}{RGB}{0,255,51}
\definecolor{ElectricGreen}{RGB}{0, 255, 0}
\definecolor{MediumPersianBlue}{RGB}{0, 103, 165}

\usepackage[unicode, colorlinks, urlcolor=BlueGreen, linkcolor=Blue, citecolor=ElectricGreen]{hyperref}

\usepackage[left=2cm,right=2cm,top=2cm,bottom=2cm,bindingoffset=0cm]{geometry}
\numberwithin{equation}{section}

\makeatletter
\newcommand{\sgn}{ \operatorname{sgn} }
\newcommand{\Tord}{ \operatorname{T} }
\newcommand{\bra}[1]{ \langle #1 | }
\newcommand{\ket}[1]{ | #1 \rangle }
\makeatother

\usepackage{authblk}

\title{\bf Loop corrections to the current of created pairs in the lengthy electric pulse}
\author[1,2]{E.~T.~Akhmedov\thanks{\href{mailto:akhmedov@itep.ru}{akhmedov@itep.ru}}}
\author[1]{P.~S.~Zavgorodny\thanks{\href{mailto:zavgorodnij.ps@phystech.edu}{zavgorodnij.ps@phystech.edu}}}
\author[1,2]{D.~I.~Sadekov\thanks{\href{mailto:sadekov.di@phystech.edu}{sadekov.di@phystech.edu}}}
\author[1]{K.~A.~Kazarnovskii\thanks{\href{mailto:kazarnovskiy.ka@phystech.edu}{kazarnovskiy.ka@phystech.edu}}}
\affil[1]{\textcolor{black}{Institutskii per, 9, Moscow Institute of Physics and Technology, 141700, Dolgoprudny, Russia}}
\affil[2]{\textcolor{black}{NRC ”Kurchatov Institute”, 123182, Moscow, Russia}}
\date{\today}

\begin{document}

\maketitle

\begin{abstract}
    We discuss loop-corrections to the electric current produced by strong and lengthy electric pulse. Namely we calculate one loop contribution to the electric current, distinguishing terms which depend on the pulse duration. We show that one loop correction does not lead to a strong modification of the tree-level current, which linearly grows with the pulse duration. Meanwhile, the correction to the photon propagator does contain the additional secular growth with the pulse duration. Based on the latter observation we argue that higher loop corrections will strongly modify the tree-level current, but that will demand a longer duration of the pulse than if the growth was at the first loop level.
\end{abstract}

\newpage

\tableofcontents

\section{Introduction}\label{sec1}
In the most of original works on the Schwinger pair creation in strong background electric fields the phenomenon was treated at tree-level \cite{Schwinger:1951nm, Grib:1976zw, Zeldovich:1971mw, Nikishov, GribBook, Grib:1980aih}. However, recent studies show that IR effects can lead to a non-negligible growing with time corrections at the loop level \cite{Akhmedov:2014hfa, Akhmedov:2014doa}, such that one should perform a resummation of loops to obtain the correct expression for the current. 

In this paper we address the issue of loop corrections to the current of created pairs in lengthy and strong electric pulse. The tree level current of created pairs is known to grow linearly with the duration of the pulse 
\cite{Gavrilov:1996pz,Kluger:1992gb,Gavrilov:2005dn,Gavrilov:2007hq,Gavrilov:2007ij,Gavrilov:2008fv,Gavrilov:2012jk,Anderson:2013zia,Anderson:2013ila} (see also \cite{Krotov:2010ma, Akhmedov:2014hfa, Akhmedov:2014doa, Akhmedov:2020dgc}). We want to understand if loop corrections, being suppressed by higher powers of the coupling constant, contain secularly growing terms depending on higher powers of the pulse duration. Then loop corrections would strongly modify the tree-level current for the lengthy enough pulse.  
In fact, the secular growth of loop corrections to the current in constant eternal electric field was first observed in \cite{Akhmedov:2014hfa,Akhmedov:2014doa}. We would like to extend this study to the case of electric pulse. The question we would like to address is if the current is growing with the length of the pulse\footnote{After the pulse, when the background electric field is switched off, we expect the thermalization process of the type considered in \cite{Akhmedov:2021vfs}. But the thermalization process after the switching off the pulse is a separate story beyond our main concern in the present paper. Here we are interested in the change of the state of the theory only during the pulse.}. 

Apart from its own interest the situation with strong electric fields in QED can be considered as the model example for the particle creation in expanding universes and other gravitational backgrounds. In fact, the same type of secular growth of loop corrections as in constant electric field \cite{Akhmedov:2014hfa,Akhmedov:2014doa} is observed e.g. in de Sitter space in \cite{Krotov:2010ma,Akhmedov:2011pj} (see \cite{Akhmedov:2013vka} and \cite{Akhmedov:2021rhq} for a reviews on the related issues), in the expanding universe of FLRW type \cite{Akhmedov:2022whm} and in the collapsing black hole backgrounds \cite{Akhmedov:2015xwa}. Of course there are other types of secular effects in de Sitter space, which have a different physical origin\footnote{Other types of infrared and/or secular effects in de Sitter-type inflationary models, such as the effective behaviour of photons and gravitons during inflation  and analytical solutions of Dyson-Schwinger equations for light fields in $O(N)$-model, have been considered e.g. in \cite{Prokopec:2007ak, Glavan:2021adm}  and \cite{Gautier:2015pca, Serreau:2013psa}, correspondingly.}
 (see \cite{Akhmedov:2019cfd} for the related discussion), but we are interested here only in those universal effects which are due to a change of the state of the theory. In fact, they are observed in many physically different backgrounds. E.g. apart from the aforementioned situations similar secular effects have been observed in strong backgrounds of other origin \cite{Akhmedov:2017hbj,Alexeev:2017ath,Akhmedov:2020haq,Trunin:2022dvg,Trunin:2021fom}. 

Also, large loop corrections in external fields are often discussed in the context of the Ritus-Narozhny conjecture. Namely the discussion of the
increase of loop contributions in a constant crossed field ($\mathbf{E}\perp\mathbf{H},\; E=H$) with field
strength can be found in \cite{Ritus1970,Narozhnyi:1980dc}  (see \cite{Fedotov:2016afw} for the recent review). The increase of loop corrections with the field
strength is expected also for the case of constant magnetic field, see for example \cite{Karbstein:2019wmj}. Finally, it is shown in \cite{Heinzl:2021mji} that the breakdown of the perturbation theory occurs even in classical electrodynamics. But all these effects are not directly related to the main subject of our paper, because our main concern here is not just a breakdown of perturbation theory. We are mostly interested in the change of the initial state of the theory and measurable consequences of such a change.

In this paper we consider the scalar QED on strong background of lengthy electric pulse and calculate one-loop correction to the electric current. On general grounds we expect that the initial state of the theory is changed under the action of the strong background field. Hence, the level population of the matter field and anomalous averages are changing in time. Namely, we are mainly interested in the dependence on the pulse duration of the level population $n_\mathbf{p}(t) \sim \langle a^+ a \rangle$ and of the anomalous quantum average $\kappa_\mathbf{p}(t) \sim \langle a b \rangle$, which do contribute to the full electric current. (It is the time dependence of these quantities, which was the reason for the origin of strong secular effects in constant electric field background \cite{Akhmedov:2014hfa,Akhmedov:2014doa} and in the gravitational backgrounds of various types, that we have mentioned above.) The question we would like to address here is if these contributions are growing rapidly enough with the duration of the pulse. 

As we consider the theory with the time-dependent free Hamiltonian, we estimate the quantum averages in question with the use of the Schwinger-Keldysh diagrammatic technique. It turns out that in the one-loop order there are no growing with the pulse duration contributions to $n_\mathbf{p}(t)$ and $\kappa_{\mathbf{p}}(t)$. As the result one loop corrections bring nothing more relevant than the UV renormalization of the tree-level expression for the current. However, correction to the photon propagator contains additional power of the pulse duration in comparison with tree-level current. That affects higher loop corrections and makes them strong. We argue that higher loop corrections do strongly modify the tree-level current.

\section{Set up of the problem}
In this paper we consider scalar electrodynamics with an external classical conserved source $j_{\mu}^{\text{cl}}, \; \partial^\mu j_{\mu}^{\text{cl}}=0$. The action is the following:
\begin{equation}\label{SQED_action_initial}
S[\phi,\phi^\dagger; A^\mu]=\int d^4x\left[|\mathcal{D}_\mu\phi|^2-m^2|\phi|^2-\dfrac{1}{4}F_{\mu\nu}F^{\mu\nu}-j_{\mu}^{\text{cl}}A^\mu \right],
\end{equation}
where $F_{\mu\nu}=\partial_\mu A_\nu-\partial_\nu A_\mu$ and $\mathcal{D}_\mu=\partial_\mu+ieA_\mu$ as usual. We choose the electric pulse as a background, which is the solution of the classical equation $\partial^\mu F_{\mu\nu}=j^{\text{cl}}_\nu$ and is homogeneous in space:
\begin{equation}\label{background_field}
A^{\text{cl}}_\mu=(0; A_1(t); 0; 0),\quad A_1(t)=ET\tanh\dfrac{t}{T}.
\end{equation}
Dividing the vector potential into the classical and quantum parts
\begin{equation}\label{quantum_A_mu}
A^\mu = A_{\text{cl}}^\mu + a^\mu,
\end{equation}
we obtain the action
\begin{equation}\label{SQED_action_finally}
S[\phi,\phi^\dagger; a^\mu]=\int d^4x\left[|D_\mu\phi + iea_{\mu}\phi|^2-m^2|\phi|^2-\dfrac{1}{4}f_{\mu\nu}f^{\mu\nu}\right],\quad
f_{\mu\nu}=\partial_\mu a_\nu-\partial_\nu a_\mu,
\end{equation}
where we denote $D_{\mu} = \partial_\mu+ieA_\mu^{\text{cl}}$ (we omit the subscript ''cl'' below for simplicity). In the following discussion we will use the Feynman gauge for $a^\mu$ and decompose the fields into the mode functions:
\begin{equation}\label{a_decomposition}
\widehat{a}_\mu(t,\mathbf{x})=\int\dfrac{d^3\mathbf{q}}{(2\pi)^3\sqrt{2|\mathbf{q}|}}
\left(
\widehat{\alpha}_{\mathbf{q}\mu}e^{-i|\mathbf{q}|t+i\mathbf{qx}}+
\widehat{\alpha}^{\dagger}_{\mathbf{q}\mu}e^{i|\mathbf{q}|t-i\mathbf{qx}}
\right),
\end{equation}
\begin{equation}\label{phi_decomposition}
\widehat{\phi}(t,\mathbf{x})=\int\dfrac{d^3\mathbf{p}}{(2\pi)^3}\left(
\widehat{a}_{\mathbf{p}}e^{i\mathbf{px}}f_{\mathbf{p}}(t)+
\widehat{b}^{\dagger}_{\mathbf{p}}e^{-i\mathbf{px}}f^{*}_{-\mathbf{p}}(t)
\right),
\end{equation}
where the creation and annihilation operators satisfy the canonical commutation relations:
\begin{equation}\label{em_KKS}
\left[
\widehat{\alpha}_{\mathbf{p}\mu},\widehat{\alpha}_{\mathbf{q}\nu}^\dagger
\right]=g_{\mu\nu}\delta(\mathbf{p}-\mathbf{q}),
\quad
\left[
\widehat{a}_{\mathbf{p}},\widehat{a}_{\mathbf{q}}^\dagger 
\right] = (2\pi)^3\delta(\mathbf{p}-\mathbf{q}),
\quad
\left[
\widehat{b}_{\mathbf{p}},\widehat{b}_{\mathbf{q}}^\dagger 
\right] = (2\pi)^3\delta(\mathbf{p}-\mathbf{q}),
\end{equation}
and the function $f_{\mathbf{p}}(t)$ solves the equation
\begin{equation}\label{modes_with_background}
\left( \partial_t^2+\left(\mathbf{p}+e\mathbf{A}(t)\right)^2+m^2\right)f_{\mathbf{p}}(t)=0.
\end{equation}
It is convenient to introduce physical $\mathbf{P}(t)=\mathbf{p}+e\mathbf{A}(t)$ and transversal $ \mathbf{p}_\perp=(p_2,p_3)$ momenta, so that $P_1(\pm\infty)=p_1\pm eET=P_{1\pm}$. The corresponding frequencies in the oscillator type equation (\ref{modes_with_background}) on the past and future infinities are $\omega_\pm(\mathbf{p})=\sqrt{\mathbf{P}^2_\pm+m^2}$. We require the harmonic function $f_{\mathbf{p}}(t)$ to be a single plane wave in the past $t \rightarrow -\infty$:
\begin{equation}\label{asympt_past}
f^{\text{in}}_\mathbf{p}(t/T\to-\infty)\simeq\dfrac{e^{-i\omega_{-}t}}{\sqrt{2\omega_{-}}}.
\end{equation}
Then the corresponding solution is referred to as the in-mode and has the following form \cite{GribBook}:
\begin{equation}\label{modes_exactly}
\begin{split}
f^{\text{in}}_{\mathbf{p}}(t) =&
\dfrac{\exp[-i\omega_{-}t]}{\sqrt{2\omega_{-}}}
\left[1+\exp\left[\dfrac{2t}{T}\right]\right]^\theta\times \\
\times &
{}_{2}F_1\left[\theta-iT\cdot\dfrac{\omega_-+\omega_+}{2},\theta-iT\cdot\dfrac{\omega_--\omega_+}{2};1-i\omega_-T;-\exp\left[\dfrac{2t}{T}\right]\right],
\end{split}
\end{equation}
where
\begin{equation}\label{theta_complex}
\theta=\dfrac{1}{2}+\sqrt{\dfrac{1}{4}-\left(eET^2\right)^2}.
\end{equation}
Below we will also use the following notations for the parameters of the hypergeometric function $ {}_{2}F_1 $ in (\ref{modes_exactly}):
\begin{equation}\label{hypergeometry_coefficients}
\xi=\theta-iT\cdot\dfrac{\omega_-+\omega_+}{2},\quad
\eta=\theta-iT\cdot\dfrac{\omega_--\omega_+}{2},\quad
\delta=1-i\omega_-T.
\end{equation}
We will also need the asymptotic form of the modes (\ref{modes_exactly}) in the future infinity $t\rightarrow +\infty$:
\begin{equation}\label{modes_asymp}
f^{\text{in}}_{\mathbf{p}}(t/T\to+\infty)\simeq
\mathcal{A}_+(\mathbf{p})e^{i\omega_+t}+\mathcal{A}_-(\mathbf{p})e^{-i\omega_+t},
\end{equation}
where
\begin{equation}\label{modes_asymp_coeffitients}
\mathcal{A}_+(\mathbf{p})=\dfrac{1}{\sqrt{2\omega_-}}\cdot\dfrac{\Gamma(\delta)\Gamma(\eta-\xi)}{\Gamma(\eta)\Gamma(\delta-\xi)},\quad \mathcal{A}_-(\mathbf{p})=\dfrac{1}{\sqrt{2\omega_-}}\cdot\dfrac{\Gamma(\delta)\Gamma(\xi-\eta)}{\Gamma(\xi)\Gamma(\delta-\eta)}.
\end{equation}
In addition, it is crucial to note here that the in-mode (\ref{modes_exactly}) leads to the proper Hadamard behaviour of the propagators. In fact, the mode function is the single plane wave for large momenta $ |\mathbf{p}| $: $f_{\mathbf{p}} \sim e^{-i|\mathbf{p}|t},\; |\mathbf{p}|\gg m,\; |\mathbf{p}|\gg eET$.

As we already mentioned in the introduction, in such a time dependent background as we consider here one has to use the Schwinger-Keldysh diagrammatic technique. In this technique gauge and scalar fields are characterized by the following propagators:
\begin{equation}\label{tree_scalar_propagator_def}
D^{ab}(t_1,\mathbf{x}_1;t_2,\mathbf{x}_2)=\bra{0}\Tord_{\mathcal{C}} \widehat{\phi}(t^a_1,\mathbf{x}_1)\widehat{\phi}^\dagger(t^b_2,\mathbf{x}_2) \ket{0},\quad a,b=+,-,
\end{equation}
\begin{equation}\label{tree_photon_propagator_def}
G^{ab}_{\mu\nu}(t_1,\mathbf{x}_1;t_2,\mathbf{x}_2)=\bra{0}\Tord_{\mathcal{C}} \widehat{a}_\mu(t^a_1,\mathbf{x}_1)\widehat{a}_\nu(t^b_2,\mathbf{x}_2) \ket{0} = g_{\mu\nu}G^{ab}(t_1,\mathbf{x}_1;t_2,\mathbf{x}_2),\quad a,b=+,-,
\end{equation}
where the $+$ and $-$ signs correspond to the upper and lower branches of the Keldysh time contour $\mathcal{C}$ and the initial state $\ket{0}$ is chosen to be the Fock space ground state, which vanishes under the action of all annihilation operators from (\ref{a_decomposition})--(\ref{phi_decomposition}); $\Tord_{\mathcal{C}}$ means the time ordering along the Keldysh time contour. 

In the calculations below we will use the propagators (\ref{tree_scalar_propagator_def})--(\ref{tree_photon_propagator_def}) in the momentum space, which is convenient due to the spatial homogeneity of the background field and the state under consideration:
\begin{equation}\label{tree_scalar_propagator_finally}
\begin{split}
D^{+-}(\mathbf{p};t_1,t_2)=f_{\mathbf{p}}(t_1)f^*_{\mathbf{p}}(t_2),\quad
D^{-+}(\mathbf{p};t_1,t_2)=f^*_{\mathbf{p}}(t_1)f_{\mathbf{p}}(t_2); \\
D^{--}(\mathbf{p};t_1,t_2)=\theta(t_1-t_2)f_{\mathbf{p}}(t_1)f^*_{\mathbf{p}}(t_2)+\theta(t_2-t_1)f^*_{\mathbf{p}}(t_1)f_{\mathbf{p}}(t_2), \\
D^{++}(\mathbf{p};t_1,t_2)=\theta(t_1-t_2)f^*_{\mathbf{p}}(t_1)f_{\mathbf{p}}(t_2)+\theta(t_2-t_1)f_{\mathbf{p}}(t_1)f^*_{\mathbf{p}}(t_2)
\end{split}
\end{equation}
and the same for the photon propagator. In terms of the notations (\ref{tree_scalar_propagator_finally}) we can write down the expression for the electric current of the matter as
\begin{equation}\label{current_via_propagator}
j_\mu(t,\mathbf{x})=ie\left[ D^{(1)}_\mu-D^{(2)\dagger}_\mu \right]D^{+-}(t_1,\mathbf{x}_1;t_2,\mathbf{x}_2)\Big{|}_{\mathbf{x}_1=\mathbf{x}_2=\mathbf{x},t_1=t_2=t},
\end{equation}
so that its first component along the electric field has the form
\begin{equation}\label{current_finally}
j_1(t)=2e\int\dfrac{d^3\mathbf{p}}{(2\pi)^3}\Bigl[p_1+eA_1(t)\Bigr]D^{+-}(\mathbf{p},t,t).
\end{equation}
Other components of the current are vanishing after the normal ordering. There is a well-known expression for the current in the long and strong pulse $eET^2\gg 1,\; eET\gg m$, which can be derived using the asymptotics (\ref{modes_asymp}) \cite{Grib:1980aih, Gavrilov:1996pz, Gavrilov:2007hq, Gavrilov:2008fv, Akhmedov:2020dgc}:
\begin{equation}\label{tree_current_renorm_finally}
j_1(t) \simeq\dfrac{E^2e^3T}{2\pi^3}\cdot\exp\left[-\dfrac{\pi m^2}{eE}\right].
\end{equation}
Our main goal in the subsequent sections is to estimate loop corrections to the expression (\ref{tree_current_renorm_finally}) starting with the eq. (\ref{current_finally}) and finding out if there is an extra growth of the corrected current on the pulse duration $T$. Namely, we want to check if higher loops, being suppressed by higher powers of the fine structure constant $e^2$, bring higher powers of the pulse duration $T$. The presence of such corrections would mean that for the very long pulse the tree-level current (\ref{tree_current_renorm_finally}) is strongly modified.

In the interaction picture creation and annihilation operators are time independent. But at the loop level the population number $ \left\langle \widehat{a}^\dagger\widehat{a}\right\rangle $ and the anomalous average $ \langle \widehat{a}\widehat{b}\rangle $ may grow with time. In view of the papers \cite{Akhmedov:2014hfa,Akhmedov:2014doa}, where such a growth of loop corrections was observed in constant and eternal electric field, there is a reason to expect such a growth of loop corrections in the pulse duration\footnote{Also such a growth is encountered in many different situations \cite{Krotov:2010ma,Akhmedov:2011pj,Akhmedov:2022whm,Akhmedov:2015xwa,Akhmedov:2019cfd,Akhmedov:2017hbj,Alexeev:2017ath,Akhmedov:2020haq,Trunin:2022dvg,Trunin:2021fom} (see \cite{Akhmedov:2013vka} and \cite{Akhmedov:2021rhq} for a reviews on the related issues).}. There the growth of the current and of the level population together with the anomalous averages was gained during the time of observation. Here we would like to see if there is such a growth during the pulse. 

\section{One-loop correction to the current}
There are two contributions to the scalar propagator at the one-loop level: the tadpole and sunset diagrams. The loop integral for the tadpole diagram contains the same UV divergence as we encounter in QED without a background field and can be removed by the appropriate counterterms, which renormalize the mass of the scalar in the bare Lagrangian (see Appendix \ref{Tadpole}). The sunset diagram (fig. \ref{image_dia_scalar}) can contribute to the growth of $ \left\langle \widehat{a}^\dagger\widehat{a}\right\rangle $ and $ \langle \widehat{a}\widehat{b}\rangle $ and can be expressed in the momentum space as follows:
\begin{figure}[h]
\center{\includegraphics[width=0.8\linewidth]{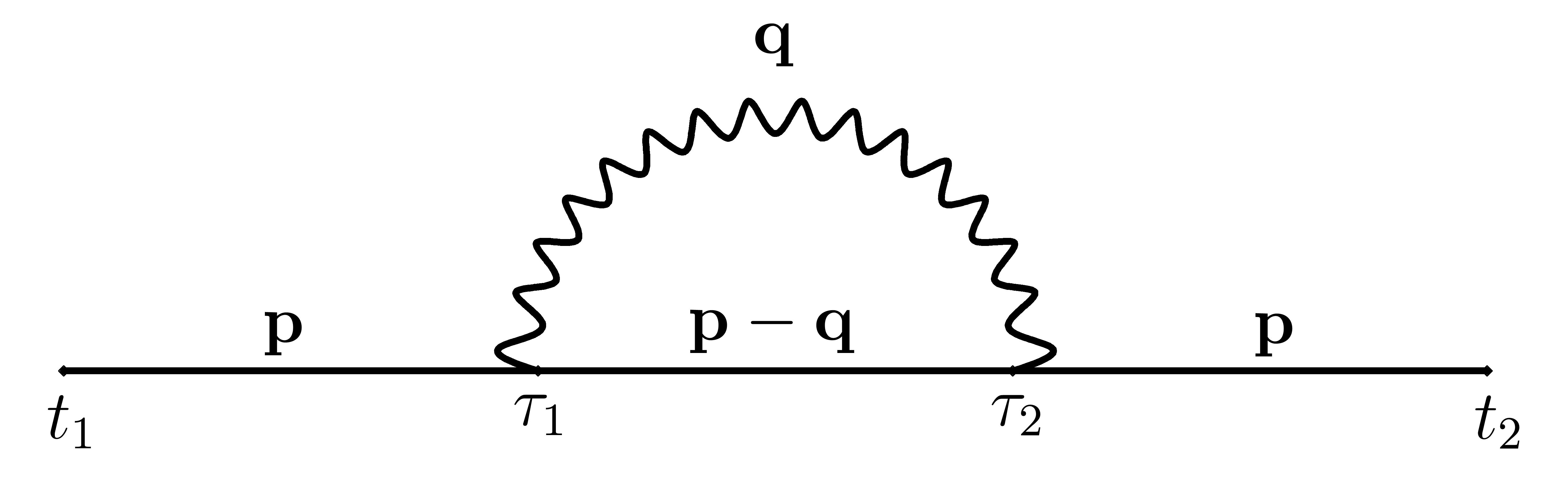}}
\caption{One-loop sunset diagram. In the Schwinger-Keldysh techniques the one loop correction contains the sum of such diagrams with different types of vertexes.}
\label{image_dia_scalar}
\end{figure}
\begin{equation}\label{loop_finally_dia}
\begin{split}
\Delta D^{ab}(\mathbf{p};t_1,t_2) &= \sum\limits_{c,d}\sgn(c)\sgn(d)e^2\int d\tau_1\int d\tau_2
\int\dfrac{d^3\mathbf{q}}{(2\pi)^3}G^{cd}(\mathbf{q},\tau_1,\tau_2)\times \\
&\times \mathbb{D}^\mu_\mu\left\{
D^{ac}(\mathbf{p},t_1,\tau_1);\;
D^{cd}(\mathbf{p}-\mathbf{q},\tau_1,\tau_2);\;
D^{db}(\mathbf{p},\tau_2,t_2)
\right\},
\end{split}
\end{equation}
where for brevity we introduce the following operator in the $\mathbf{x}$--space
\begin{equation}\label{D_big_def}
\mathbb{D}^{\mu\nu}=\left[D^\mu\right]^\dagger_{1}\left[D^\nu\right]^\dagger_{2}-\left[D^\mu\right]^\dagger_{1}D^\nu_2-D^\mu_1\left[D^\nu\right]^\dagger_{2}+D^\mu_1 D^\nu_2,
\end{equation}
where the indices $\{1,2\}$ denote the action on the corresponding coordinates of internal vertices respectively. Then the derivatives without the dagger correspond to the action on the first argument of the propagators, while the daggered derivatives act on the second argument of the propagators, i.e. on the fields $\phi$ and $\phi^{\dagger}$ correspondingly. Hence inside the expression (\ref{loop_finally_dia}) in the momentum space one obtains:
\begin{equation}\label{D_mumu_finally}
\mathbb{D}=\mathbb{D}^\mu_\mu=\mathbb{D}^0_0+\left(2\mathbf{p}-\mathbf{q}+2e\mathbf{A}(\tau_1)\right)\left(2\mathbf{p}-\mathbf{q}+2e\mathbf{A}(\tau_2)\right).
\end{equation}
Finally, substituting the equations (\ref{tree_scalar_propagator_finally}) and (\ref{loop_finally_dia}) into the (\ref{current_finally}) we get the following expression for the one-loop correction to the current:
\begin{equation}\label{loop_current_finally}
j_1(t)\approx 2e\int\dfrac{d^3\mathbf{p}}{(2\pi)^3}\left(p_1+eA_1(t)\right)\left[
\left|f_\mathbf{p}(t)\right|^2 n_\mathbf{p}(t)+
\left(f^*_\mathbf{p}(t)\right)^2 \kappa_\mathbf{p}(t)+
\left(f_\mathbf{p}(t)\right)^2 \kappa^*_\mathbf{p}(t)
\right],
\end{equation}
where
\begin{equation}\label{n-p(t)}
n_\mathbf{p}(t)\approx -2e^2\int\limits_{-\infty}^{t}d\tau_1\int\limits_{-\infty}^{t}d\tau_2\int\dfrac{d^3\mathbf{q}}{(2\pi)^3}\dfrac{e^{i|\mathbf{q}|(\tau_1-\tau_2)}}{2|\mathbf{q}|}\cdot
\mathbb{D}\bigg(
f^*_\mathbf{p}(\tau_1);\;
f^*_{\mathbf{p}-\mathbf{q}}(\tau_1)
f_{\mathbf{p}-\mathbf{q}}(\tau_2);\;
f_\mathbf{p}(\tau_2)
\bigg)
\end{equation}
and
\begin{equation}\label{kappa-p(t)}
\kappa_\mathbf{p}(t)\approx 2e^2\int\limits_{-\infty}^{t}d\tau_1\int\limits_{-\infty}^{\tau_1}d\tau_2\int\dfrac{d^3\mathbf{q}}{(2\pi)^3}\dfrac{e^{i|\mathbf{q}|(\tau_1-\tau_2)}}{2|\mathbf{q}|}\cdot
\mathbb{D}\bigg(
f_\mathbf{p}(\tau_1);\;
f^*_{\mathbf{p}-\mathbf{q}}(\tau_1)
f_{\mathbf{p}-\mathbf{q}}(\tau_2);\;
f_\mathbf{p}(\tau_2)
\bigg).
\end{equation}
The quantities $n_\mathbf{p}(t), \kappa_\mathbf{p}(t)$ are the above-mentioned level population $ \left\langle \widehat{a}^\dagger\widehat{a}\right\rangle $ and anomalous average $ \langle \widehat{a}\widehat{b}\rangle $ for the charged scalar field.

\subsection{The contribution of $n_\mathbf{p}(t)$}
In this subsection we calculate the contribution of the level population to the one-loop corrected current:
\begin{equation}\label{non_anomaly}
j^{(n)}_1(t)=2e\int\dfrac{d^3\mathbf{p}}{(2\pi)^3}\Big[p_1+eA_1(t)\Big]\cdot
\left|f_\mathbf{p}(t)\right|^2 n_\mathbf{p}(t).
\end{equation}
For $ \tau<-T $, the background electric field is not yet turned on and the modes $ f_\mathbf{p}(\tau) $ are single waves. Initial state is the Fock space ground state and the quantum averages $ \left\langle \widehat{a}^\dagger\widehat{a}\right\rangle $ and $ \langle \widehat{a}\widehat{b}\rangle $ do not yet evolve in time. For $ \tau>T $, the electric field is switched off, but the modes $ f_\mathbf{p}(\tau) $ are linear combinations of single waves, (\ref{modes_asymp}). As we show in the appendix \ref{Time_regions_outside-n}, after the pulse there are no growing contributions to  $ \left\langle \widehat{a}^\dagger\widehat{a}\right\rangle $. However, there can be a non-trivial thermalization process due to generated during the pulse $ \left\langle \widehat{a}^\dagger\widehat{a}\right\rangle $ and $ \langle \widehat{a}\widehat{b}\rangle $. To move further in this direction it is convenient to perform the Bogolubov transformation to the out-modes (single waves in the future infinity):
\begin{equation}
\widehat{c}_\mathbf{p}=\sqrt{2\omega_+}\left(
\mathcal{A}_-(\mathbf{p})\widehat{a}_\mathbf{p}+
\mathcal{A}^*_+(\mathbf{p})\widehat{b}^\dagger_{-\mathbf{p}}
\right);\quad
\widehat{d}^\dagger_{-\mathbf{p}}=\sqrt{2\omega_+}\left(
\mathcal{A}_+(\mathbf{p})\widehat{a}_\mathbf{p}+
\mathcal{A}^*_-(\mathbf{p})\widehat{b}^\dagger_{-\mathbf{p}}
\right)
\end{equation}
and
\begin{equation}
\widehat{\phi}(\tau,\mathbf{x})\simeq\int\dfrac{d^3\mathbf{p}}{(2\pi)^3}\left(
\widehat{c}_\mathbf{p}\dfrac{e^{-i\omega_+\tau+i\mathbf{px}}}{\sqrt{2\omega_+}}+
\widehat{d}^\dagger_\mathbf{p}\dfrac{e^{i\omega_-\tau-i\mathbf{px}}}{\sqrt{2\omega_-}}
\right)
\end{equation}
Similar thermalization process was considered in \cite{Akhmedov:2021vfs} for the $ \lambda\varphi^4 $-theory. But that is a separate story going beyond our main interest in the present paper. 

Thus, interesting for us contribution to the current can be approximated by:
\begin{equation}\label{j^n_1(t)}
j^{(n)}_1(t)\simeq 2e\int\dfrac{d^3\mathbf{p}}{(2\pi)^3}\left(p_1+eA_1(t)\right)\cdot
\left|f_\mathbf{p}(t)\right|^2 \mathcal{N}_\mathbf{p}(T),
\end{equation}
where $ \mathcal{N}_\mathbf{p}(T) $ differs from $ n_\mathbf{p}(t) $ in that the integration over $ \tau_{1,2} $ in the former is restricted to the domain $ |\tau_{1,2}|<T $:
\begin{equation}\label{N_p(T)}
\mathcal{N}_\mathbf{p}(T)=-2e^2\int\limits_{-T}^{T}d\tau_1\int\limits_{-T}^{T}d\tau_2\int\dfrac{d^3\mathbf{q}}{(2\pi)^3}\dfrac{e^{i|\mathbf{q}|(\tau_1-\tau_2)}}{2|\mathbf{q}|}\cdot
\mathbb{D}\bigg(
f^*_\mathbf{p}(\tau_1);\;
f^*_{\mathbf{p}-\mathbf{q}}(\tau_1)
f_{\mathbf{p}-\mathbf{q}}(\tau_2);\;
f_\mathbf{p}(\tau_2)
\bigg).
\end{equation}
To calculate this expression we have to investigate the behaviour of the in-mode $ f_\mathbf{p}(t) $ in the region $ |t|<T $. The hypergeometric function $ {}_{2}F_1 $ in (\ref{modes_exactly}) is difficult to work with in the region $ |t|<T $, so we will use the approximation of the pulse with the constant electric field in this region and find the approximate form of the modes there. The corresponding differential equation
\begin{equation}\label{modes_with_background_eternal}
\left( \partial_\tau^2+
\left(p_1+eE\tau\right)^2+
\mathbf{p}^2_\perp+
m^2\right)f_{\mathbf{p}}(\tau)=0
\end{equation}
has a general solution
\begin{equation}\label{straight_modes}
\begin{split}
f_{\mathbf{p}}(t) &= g_{\mathbf{p}_\perp}(p_1+eEt)= \\
&= c_1 D_{-\frac{1}{2}+i\frac{\mathbf{p}^2_\perp+m^2}{2eE}}
\left[ \dfrac{(1-i)(p_1+eEt)}{\sqrt{eE}} \right]+
c_2 D_{-\frac{1}{2}+i\frac{\mathbf{p}^2_\perp+m^2}{2eE}}
\left[-\dfrac{(1-i)(p_1+eEt)}{\sqrt{eE}} \right],
\end{split}
\end{equation}
where $D_{\nu}(z)$ is the parabolic cylinder function and the coefficients are some functions of momenta $c_{1,2} = c_{1,2}(\mathbf{p})$, which should be chosen in such a way as to approximate the exact in-mode (\ref{modes_exactly}) with (\ref{straight_modes}) in the region $ |t|<T $. The integral over $p_1$ in (\ref{j^n_1(t)}) is saturated within the region $|p_1|<eET$ as for large momenta the mode reduces to the single wave. Therefore, we equate the solutions (\ref{straight_modes}) and (\ref{modes_exactly}) and its derivatives around the point $p_1 = 0,\;t=0$, to obtain that:
\begin{equation}\label{C_1C_2}
\begin{split}
c_1+c_2 &= \dfrac{2^{ieET^2}}{2^{\xi_0/2}\sqrt{eET}}\cdot\dfrac{\Gamma(\delta_0)}{\Gamma\left[\dfrac{\xi_0}{2}+1-\eta_0\right]}, \\
c_1-c_2 &= \dfrac{(\eta_0-\delta_0)2^{ieET^2}}{2^{\xi_0/2+1}e^{-i\pi/4}eET\sqrt{T}}\cdot\dfrac{\Gamma(\delta_0)}{\Gamma\left[\dfrac{\xi_0+3}{2}-\eta_0\right]},
\end{split}
\end{equation}
where $ \xi_0, \eta_0, \delta_0 $ are respectively equal to $ \xi, \eta, \delta $ from (\ref{hypergeometry_coefficients}) at $ p_1=0 $.

In order to establish the dependence of the coefficients on $T$ we expand all the parameters in (\ref{hypergeometry_coefficients}) in powers of $\frac{1}{T}$ and use the asymptotics of the Gamma function
\begin{equation}
\ln\Gamma(z)=\left[z-\dfrac{1}{2}\right]\ln z-z+\dfrac{\ln(2\pi)}{2}+O\left[\dfrac{1}{z}\right],\quad |\mathrm{arg} z|<\pi
\end{equation}
to find that
\begin{equation}\label{Gamma-estimation}
\dfrac{\Gamma(\delta_0)}{\Gamma\left[\dfrac{\xi_0+\beta_\pm}{2}-\eta_0\right]}=\left[eET^2\right]^{\pm 1/4}
\left[eET^2\right]^{-i\frac{\mathbf{p}^2_\perp+m^2}{4eE}}\cdot
e^{\mp i\pi/8}\cdot\exp\left[-\dfrac{\pi(\mathbf{p}^2_\perp+m^2)}{8eE} \right]\cdot\left[1+\mathcal{O}\left(\dfrac{1}{T^2}\right)\right],
\end{equation}
where $\beta_{+} = 2,\;\beta_{-}=3$. Hence, the coefficients under consideration are as follows:
\begin{equation}\label{cylinder_coefs}
c\equiv c_2\simeq\dfrac{2^{ieET^2}}{2^{\xi_0/2}[eE]^{1/4}}
\left[eET^2\right]^{-i\frac{\mathbf{p}^2_\perp+m^2}{4eE}}\cdot
e^{-i\pi/8}\cdot\exp\left[-\dfrac{\pi(\mathbf{p}^2_\perp+m^2)}{8eE} \right],\quad c_1\simeq 0,
\end{equation}
\begin{equation}\label{cylinder_coefs_abs}
\left|c(\mathbf{p}_\perp)\right|^2\simeq\dfrac{1}{\sqrt{2eE}}\exp\left[-\dfrac{\pi(\mathbf{p}^2_\perp+m^2)}{4eE}\right],
\end{equation}
up to the order of $\mathcal{O}\left(\dfrac{1}{T^2}\right)$. Finally, let us introduce the dimensionless variables
\begin{gather}\label{dimensionless_vars}
\chi=\dfrac{p_1+eET}{\sqrt{eE}},\quad
\chi_i=\dfrac{p_1+eE\tau_i}{\sqrt{eE}},\quad
\mathbf{Q}=\dfrac{\mathbf{q}}{\sqrt{eE}},
\nonumber
\\
X=\dfrac{\chi_1+\chi_2}{2},\quad \widetilde{\chi}=\chi_1-\chi_2,
\end{gather}
such that the integral (\ref{N_p(T)}) acquires the form
\begin{equation}\label{N_p(T)-2}
\begin{split}
\mathcal{N}_{\mathbf{p}_\perp}\left(\chi\right)\approx -2e^2
\int\limits_{\chi-2\sqrt{eE}T}^{\chi}dX
\int\limits_{-\infty}^{+\infty}d\widetilde{\chi}
\int\dfrac{d^3\mathbf{Q}}{(2\pi)^3}\dfrac{e^{i|\mathbf{Q}|\widetilde{\chi}}}{2|\mathbf{Q}|}\times \\
\times \mathbb{D}\bigg(
g^*_{\mathbf{p}_\perp}(X+\widetilde{\chi}/2);\;
g^*_{\mathbf{p}_\perp-\mathbf{q}_\perp}(X+\widetilde{\chi}/2-Q_1)
g_{\mathbf{p}_\perp-\mathbf{q}_\perp}(X-\widetilde{\chi}/2-Q_1);\;
g_{\mathbf{p}_\perp}(X-\widetilde{\chi}/2)
\bigg).
\end{split}
\end{equation}
As we are concerned only about the dependence of $ \mathcal{N}_{\mathbf{p}_\perp}(\chi) $ on $T$, we can represent the integral over $ X $ in (\ref{N_p(T)-2}) as the difference between the integral $\mathcal{N}_{\mathbf{p}_\perp}^\infty$ over the whole real axis $ X\in(-\infty;+\infty) $ and the integral over the redundant domain. Since $\mathcal{N}_{\mathbf{p}_\perp}^\infty$ does not depend on $ T $, only the integration over redundant domain is of interest. Also due to the Hadamard behaviour of the exact harmonics we can limit ourselves to the region $|p_1|<\left(1-\delta\right)eET, \; \delta \lesssim \frac{1}{\sqrt{eE}T}$ (the narrow area around $p_1=eET$ doesn't bring anything qualitatively new) and note that the integral over $\mathbf{Q}$ in (\ref{N_p(T)-2}) is saturated at $|\widetilde{\chi}|\sim 1 $. Then we can apply the asymptotics of the parabolic cylinder functions for large argument in these domains \cite{Bateman:100233}:
\begin{equation}\label{cylinder_asymptotics-1}
g_{\mathbf{p}_\perp}(X-\widetilde{\chi}/2)\simeq
\alpha_\pm(\mathbf{p}_\perp)\dfrac{(X-\widetilde{\chi}/2)^{i\frac{m^2+\mathbf{p}^2_\perp}{2eE}}}{(X-\widetilde{\chi}/2)^{1/2}}
e^{\frac{i(X-\widetilde{\chi}/2)^2}{2}}+
\beta_\pm(\mathbf{p}_\perp)\dfrac{(X-\widetilde{\chi}/2)^{-i\frac{m^2+\mathbf{p}^2_\perp}{2eE}}}{(X-\widetilde{\chi}/2)^{1/2}}
e^{\frac{-i(X-\widetilde{\chi}/2)^2}{2}},
\end{equation}
where the sign $ \pm $ corresponds to the sign of $X$ and 
\begin{equation}\label{coefs_in_D}
\begin{split}
\left|\alpha_\pm(\mathbf{p}_\perp)\right|^2 &= \dfrac{1}{2\sqrt{eE}}\exp\left(-\dfrac{\pi(\mathbf{p}^2_\perp+m^2)}{eE}\right) \\
\left|\beta_+(\mathbf{p}_\perp)\right|^2 &= \dfrac{1}{2\sqrt{eE}}+\left|\alpha_\pm(\mathbf{p}_\perp)\right|^2, \quad \left|\beta_-(\mathbf{p}_\perp)\right|^2 = 0.
\end{split}
\end{equation}
Finally, the estimation of (\ref{N_p(T)-2}) is given by
\begin{equation}
\mathcal{N}_{\mathbf{p}_\perp}\left(\chi\right)\simeq
\mathcal{N}_{\mathbf{p}_\perp}^\infty-2e^2
\int\limits_{\chi}^{+\infty}\dfrac{dX}{X^2}\cdot
h(\mathbf{p}_\perp)= \mathcal{N}_{\mathbf{p}_\perp}^\infty-
\dfrac{2e^2h(\mathbf{p}_\perp)}{p_1+eET},
\end{equation}
where
\begin{equation}
\begin{split}
h(\mathbf{p}_\perp)&=-\int\limits_{-\infty}^{+\infty}d\widetilde{\chi}
\int\dfrac{d^3\mathbf{Q}}{(2\pi)^3}\dfrac{e^{i|\mathbf{Q}|\widetilde{\chi}}}{2|\mathbf{Q}|}\left( 2\mathbf{p}_\perp-\mathbf{q}_\perp \right)^2\times \\
&\times\left[|\alpha_+(\mathbf{p}_\perp)|^2|\beta_+(\mathbf{p}_\perp-\mathbf{q}_\perp)|^2e^{-iQ_1\widetilde{\chi}}+
|\beta_+(\mathbf{p}_\perp)|^2|\alpha_+(\mathbf{p}_\perp-\mathbf{q}_\perp)|^2e^{iQ_1\widetilde{\chi}}\right].
\end{split}
\end{equation}
Here we kept only the non-suppressed oscillating terms in the limit $|\chi|\gg 1$ and the leading order in the expansion in powers of $\frac{1}{T}$. Hence, the contribution (\ref{non_anomaly}) to the current can be estimated as:
\begin{equation}\label{non_anomaly_final}
j^{(n)}_1(t)\simeq 2e^2ET
\int\dfrac{d^2\mathbf{p}_\perp}{(2\pi)^3}
\left[2\exp\left(-\dfrac{\pi(\mathbf{p}^2_\perp+m^2)}{eE}\right)+1\right]
\mathcal{N}_{\mathbf{p}_\perp}^\infty.
\end{equation}
One can see that (\ref{non_anomaly_final}) doesn't bring an additional positive power of $T$ in comparison with the classical result (\ref{tree_current_renorm_finally}). 

\subsection{The contribution of $\kappa_\mathbf{p}(t)$}
In this subsection we calculate the contribution of the anomalous average to the one-loop corrected current:
\begin{equation}\label{anomalous_current_initial}
j^{(\kappa)}_1(t)=2e\int\dfrac{d^3\mathbf{p}}{(2\pi)^3}\left(p_1+eA_1(t)\right)\left[
\left(f^*_\mathbf{p}(t)\right)^2 \kappa_\mathbf{p}(t)+
\text{h.c.}\right].
\end{equation}
As we show in the appendix \ref{Time_regions_outside-kappa}, anomalous average $ \kappa_\mathbf{p}(t) $ grows with $ t $, in contrast to $ n_\mathbf{p}(t) $. However, this growth is canceled\footnote{The growth of the anomalous average has a clear physical meaning -- the initial state that we consider (the Fock space ground state) is not going to be the final state of the theory. Furthermore, despite the fact that anomalous average does not contribute growing with $t$ contribution to the current, there can be growing with $t$ loop corrections to the stress energy tensor. And, as we have already mentioned above, there is a certain interesting physics after the pulse is switched off, which is related to the thermalization process.} after the substitution of $ \kappa_\mathbf{p}(t) $ into (\ref{anomalous_current_initial}). Thus, the time region after the pulse does not bring loop corrections into the current comparable to the tree-level contribution (\ref{tree_current_renorm_finally}). Hence, we consider only the region $ -T<\tau_2<\tau_1<T $, i.e. during the pulse. The corresponding contribution to the current can be approximated by:
\begin{equation}\label{j^k_1(t)}
j^{(\kappa)}_1(T)\simeq 2e\int\dfrac{d^3\mathbf{p}}{(2\pi)^3}\left(p_1+eA_1(t)\right)\cdot
\left[\left(f^*_\mathbf{p}(t)\right)^2 \mathcal{K}_\mathbf{p}(T)+\text{h.c.}
\right],
\end{equation}
where $ \mathcal{K}_\mathbf{p}(T) $ differs from $ \kappa_\mathbf{p}(t) $ in that the integration over $ \tau_{1,2} $ in the former is restricted to the domain $ -T<\tau_2<\tau_1<T $:
\begin{equation}\label{Kappa_1}
\mathcal{K}_\mathbf{p}(T)=2e^2\int\limits_{-T}^{T}d\tau_1\int\limits_{-T}^{\tau_1}d\tau_2\int\dfrac{d^3\mathbf{q}}{(2\pi)^3}\dfrac{e^{i|\mathbf{q}|(\tau_1-\tau_2)}}{2|\mathbf{q}|}\cdot
\mathbb{D}\bigg(
f_\mathbf{p}(\tau_1);\;
f^*_{\mathbf{p}-\mathbf{q}}(\tau_1)
f_{\mathbf{p}-\mathbf{q}}(\tau_2);\;
f_\mathbf{p}(\tau_2)
\bigg).
\end{equation}
Next step is analogous to the one in the calculation of  $\mathcal{N}_\mathbf{p}(T)$. Let us use the approximate form of the modes (\ref{straight_modes}) and the change of variables (\ref{dimensionless_vars}) to write (\ref{Kappa_1}) in the following form:

\begin{equation}\label{K_p(T)-2}
\begin{split}
\mathcal{K}_\mathbf{p}(T)\equiv\mathcal{K}_{\mathbf{p}_\perp}\left(\chi\right)=2e^2
\int\limits_{\chi-2\sqrt{eE}T}^{\chi}dX
\int\limits_{0}^{+\infty}d\widetilde{\chi}
\int\dfrac{d^3\mathbf{Q}}{(2\pi)^3}\dfrac{e^{i|\mathbf{Q}|\widetilde{\chi}}}{2|\mathbf{Q}|}\times \\
\times \mathbb{D}\left[
g_{\mathbf{p}_\perp}(X+\widetilde{\chi}/2);
g^*_{\mathbf{p}_\perp-\mathbf{q}_\perp}(X+\widetilde{\chi}/2-Q_1)
g_{\mathbf{p}_\perp-\mathbf{q}_\perp}(X-\widetilde{\chi}/2-Q_1);
g_{\mathbf{p}_\perp}(X-\widetilde{\chi}/2)
\right].
\end{split}
\end{equation}
Unlike the contribution of the level population, expression (\ref{K_p(T)-2}) contains the squared coefficient $ c(\mathbf{p}_\perp) $ itself rather than its squared modulus. Even though the phase of $\mathcal{K}_{\mathbf{p}_\perp}\left(\chi\right)$ can depend on $T$, we will show further that this does not bring additional positive powers of $T$ into the current.

In the same way as in the previous section we extend the integration over $X$ and subtract the redundant domains of integration, where we use the asymptotics  (\ref{cylinder_asymptotics-1}) to obtain:
\begin{equation}
\mathcal{K}_{\mathbf{p}_\perp}(\chi)\simeq\mathcal{K}_{\mathbf{p}_\perp}^\infty-2e^2\int\limits_{\chi}^{+\infty}\dfrac{dX}{X^2}\cdot l(\mathbf{p}_\perp)=\mathcal{K}_{\mathbf{p}_\perp}^\infty-\dfrac{2e^2l(\mathbf{p}_\perp)}{p_1+eET},
\end{equation}
where
\begin{equation}
\begin{split}
l(\mathbf{p}_\perp)=\int\limits_{0}^{+\infty}dx\int\dfrac{d^3\mathbf{Q}}{(2\pi)^3}\cdot\dfrac{e^{i|\mathbf{Q}|x}}{2|\mathbf{Q}|}(2\mathbf{p}_\perp-\mathbf{q}_\perp)^2\times \\
\times\Big{[}
\alpha_+(\mathbf{p}_\perp)\beta_+(\mathbf{p}_\perp)|\alpha_+(\mathbf{p}_\perp-\mathbf{q}_\perp)|e^{iQ_1x}+
\alpha_+(\mathbf{p}_\perp)\beta_+(\mathbf{p}_\perp)|\beta_+(\mathbf{p}_\perp-\mathbf{q}_\perp)|e^{-iQ_1x}
\Big{]}.
\end{split}
\end{equation}
The corresponding contribution to the current is as follows:

\begin{equation}\label{anomalous_current}
j^{(\kappa)}_1(T)\simeq 2e^2ET\int\dfrac{d^2\mathbf{p}_\perp}{(2\pi)^3}\cdot eET\cdot\left[ 2\mathcal{A}^*_+(\mathbf{p})\mathcal{A}^*_-(\mathbf{p})\mathcal{K}^\infty_{\mathbf{p}_\perp}+\text{h.c.}\right].
\end{equation}

As we have mentioned above the phase of $\mathcal{K}_{\mathbf{p}_\perp}^\infty$ depends on $T$, so let us consider the integral over $\mathbf{p}_\perp$ in greater details. In the approximation  $ p_1\simeq 0 $ we can write

\begin{equation}
\begin{split}
\mathcal{A}_+(\mathbf{p})\mathcal{A}_-(\mathbf{p})\Big{|}_{p_1=0} =& 
\dfrac{1}{2\omega_0}\dfrac{\Gamma^2(\delta_0)\Gamma(\eta_0-\xi_0)\Gamma(\xi_0-\eta_0)}{\Gamma(\eta_0)\Gamma(\delta_0-\xi_0)\Gamma(\delta_0-\eta_0)\Gamma(\xi_0)} \\
=& \dfrac{1}{2\omega_0}\dfrac{\Gamma^2(1-i\omega_0T)|\Gamma(i\omega_0T)|^2}
{\left|\Gamma\left[\dfrac{1}{2}+i\beta\right]\right|^2\Gamma(\xi_0)
\Gamma\left[\dfrac{1}{2}-i(\beta+\omega_0T)\right]},
\end{split}
\end{equation}
and use the expression for the square of the modulus of Gamma functions:

\begin{equation}
\dfrac{\left|\Gamma(i\omega_0T)\right|^2}{\left|\Gamma\left[\dfrac{1}{2}+i\beta\right]\right|^2}\simeq\dfrac{1}{eET^2}\exp\left[-\dfrac{\pi(m^2+\mathbf{p}_\perp^2)}{2eE}\right].
\end{equation}
For other multipliers, which depend on $T$, we apply the asymptotics
\begin{equation}
    \ln\Gamma(x-iy)=\left[x-iy-\dfrac{1}{2}\right]\ln y-\dfrac{i\pi}{2}\left[x-\dfrac{1}{2}\right]-\dfrac{y\pi}{2}+iy+\dfrac{\ln(2\pi)}{2}+O\left[\dfrac{1}{y}\right],\ x\ll y,\  1\ll y
\end{equation}
to obtain:

\begin{equation}
\dfrac{\Gamma^2(1-i\omega_0T)}{\Gamma\left[\dfrac{1}{2}-i(\beta+\omega_0T)\right]}=\sqrt{2\pi}\left[eET^2\right]^{1-i\frac{m^2+\mathbf{p}_\perp^2}{2eE}}\cdot 2^{2ieET^2+i\frac{m^2+\mathbf{p}_\perp^2}{2eE}}\cdot
e^{-\frac{i\pi}{2}-\frac{\pi(m^2+\mathbf{p}_\perp^2)}{4eE}}
\end{equation}
Getting all together with the expression for $ c(\mathbf{p}_\perp) $ (\ref{cylinder_coefs}) we obtain that $ eET\cdot\mathcal{A}^*_+(\mathbf{p})\mathcal{A}^*_-(\mathbf{p})c^2(\mathbf{p}_\perp) $ is independent of $T$. Therefore the corresponding contribution to the current (\ref{anomalous_current}) depends on $T$ only via the coefficient $ e^2ET $ before the integral, thus it also does not give an additional growth in $T$. 

In all, the one loop contributions of the level population and anomalous average do not lead to any additional growth of the current $ T $. And the relevant effect of the one loop correction is just the UV renormalization of the theory.

\section{One-loop correction to the photon's Keldysh propagator}

In this section we calculate the correction to the photon's Keldysh propagator. Despite the fact that such a quantity has no effect on the current at one-loop level, as we will show below it can be relevant for the higher loop corrections. On the other hand, the quantum dynamics of the photon state in such a situation presents a separate interest itself. That is the reason why we consider Keldysh propagator and ignore advanced or retarded. The latter two on the tree level are sensitive only to the spectrum of the theory rather than the state and, hence, do not contribute to the current directly.
\begin{figure}[h]
\center{\includegraphics[width=0.6\linewidth]{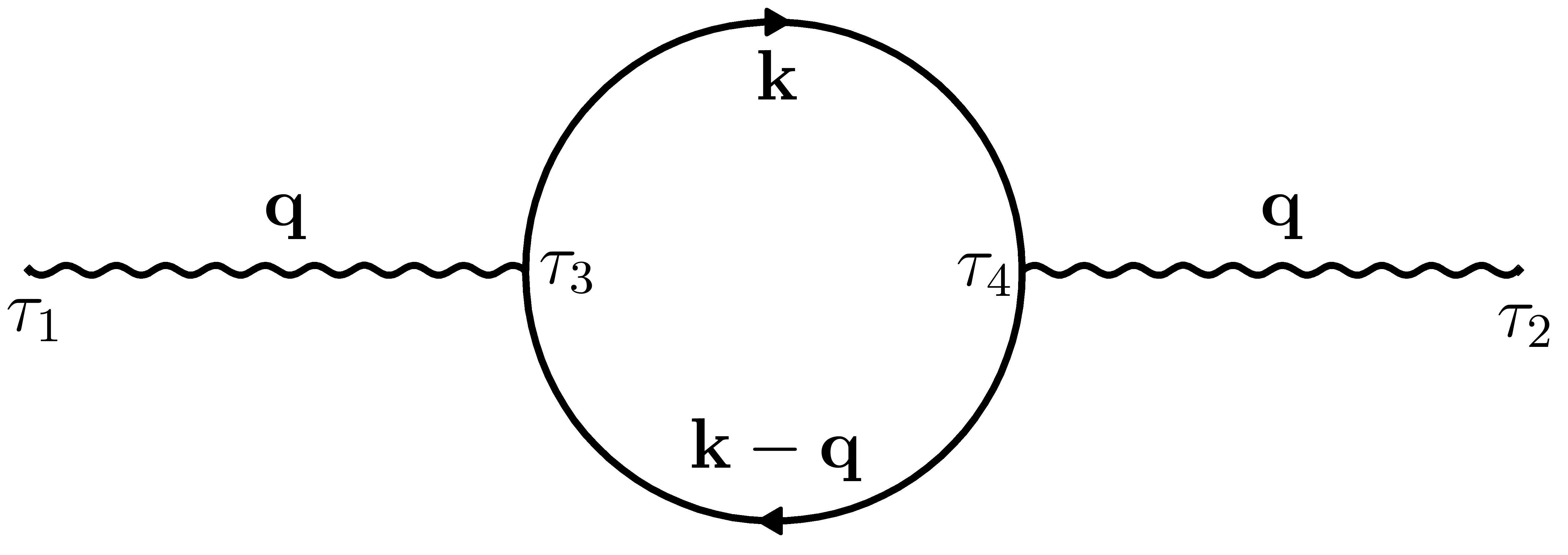}}
\caption{One-loop correction to the photon's propagator. In the Schwinger-Keldysh technique one has to sum over $\pm$ signs in the internal and external vertices}
\label{image_dia_photon}
\end{figure}

The expression for the diagram shown on the fig. \ref{image_dia_photon} takes the form:
\begin{equation}\label{photon_prop_loop_general_p}
\begin{split}
G^{ab}_{\mu\nu,\text{loop}}(\mathbf{q};\tau_1,\tau_2) =& e^2\sum\limits_{cd}\sgn(c)\sgn(d)\int d\tau_3 d\tau_4 \int\dfrac{d^3\mathbf{k}}{(2\pi)^3} \times \\
\times&
G^{ac}(\mathbf{q};\tau_1,\tau_3)
G^{db}(\mathbf{q};\tau_4,\tau_2)\times \\
\times& \mathbb{D}_{\mu\nu}\left[
D^{cd}(\mathbf{k};\tau_3,\tau_4)
D^{dc}(\mathbf{k}-\mathbf{q},\tau_4,\tau_3)
\right].
\end{split}
\end{equation}
Then the Keldysh propagator in the approximation $ |\tau_1-\tau_2|\ll|\tau_1+\tau_2| $, which is appropriate for the further calculation of the two-loop correction to the current is as follows:
\begin{equation}\label{K_photon}
G^K_{\mu\nu}(\mathbf{q};\tau_1,\tau_2)\simeq n_{\mu\nu}(\mathbf{q},\mathcal{T})\dfrac{e^{-i|\mathbf{q}|(\tau_1-\tau_2)}}{2|\mathbf{q}|}+\kappa_{\mu\nu}(\mathbf{q},\mathcal{T})\dfrac{e^{-i2|\mathbf{q}|\mathcal{T}}}{2|\mathbf{q}|}+\text{h.c.},
\end{equation}
where $ \mathcal{T}=(\tau_1+\tau_2)/2 $ and
\begin{equation}\label{n_munu}
n_{\mu\nu}(\mathbf{q},\mathcal{T})\simeq -e^2\int\limits_{t_0}^{\mathcal{T}}d\tau_3\int\limits_{t_0}^{\mathcal{T}}d\tau_4
\dfrac{e^{-i|\mathbf{q}|(\tau_3-\tau_4)}}{2|\mathbf{q}|}\int\dfrac{d^3\mathbf{k}}{(2\pi)^3}\mathbb{D}_{\mu\nu}\left[
f_{\mathbf{k}}(\tau_3)f^*_{\mathbf{k}}(\tau_4);
f^*_{\mathbf{k}-\mathbf{q}}(\tau_4)f_{\mathbf{k}-\mathbf{q}}(\tau_3)
\right],
\end{equation}
\begin{equation}\label{kappa_munu}
\kappa_{\mu\nu}(\mathbf{q},\mathcal{T})\simeq 2e^2\int\limits_{t_0}^{\mathcal{T}}d\tau_3\int\limits_{t_0}^{\tau_3}d\tau_4
\dfrac{e^{i|\mathbf{q}|(\tau_3+\tau_4)}}{2|\mathbf{q}|}\int\dfrac{d^3\mathbf{k}}{(2\pi)^3}\mathbb{D}_{\mu\nu}\left[
f_{\mathbf{k}}(\tau_3)f^*_{\mathbf{k}}(\tau_4);
f^*_{\mathbf{k}-\mathbf{q}}(\tau_4)f_{\mathbf{k}-\mathbf{q}}(\tau_3)
\right].
\end{equation}
We do not expect secular growth of $\kappa_{\mu\nu}(\mathbf{q},\mathcal{T})$ (\ref{kappa_munu}), because the interaction doesn't change the initial Poincare invariant vacuum state of photons. One can explicitly see that $ \kappa_{\nu\nu}(\mathbf{q},\mathcal{T}) $ does not grow with $ \mathcal{T} $. For this reason we put down the calculation of $\kappa_{\mu\nu}(\mathbf{q},\mathcal{T})$ and concentrate on the most interesting part -- $n_{\mu\nu}(\mathbf{q},\mathcal{T})$ (\ref{n_munu}).  

In order to analyse the expression for $n_{\mu\nu}(\mathbf{q},\mathcal{T})$ let us make the following change of variables:
\begin{equation}
\mathcal{T}_{34}=\dfrac{\tau_3+\tau_4}{2},\quad\tau_{34}=\dfrac{\tau_3-\tau_4}{2}
\end{equation}
As we have oscillating exponent in the integrand in (\ref{n_munu}) the integral over $\tau_{34}$ is rapidly saturated. Hence, we can expand the integration over $\tau_{34}$ to the entire real axis. Next, in the region $\mathcal{T}_{34}<-T$ the modes $f$ in (\ref{n_munu}) have the asymptotic form of single waves. Hence, the integral over $\tau_{34}$ gives delta-functions with never-vanishing arguments. Then we can write:
\begin{equation}\label{n_munu2}
n_{\mu\nu}(\mathbf{q},\mathcal{T})\simeq -2e^2\int\limits_{-T}^{\mathcal{T}}d\mathcal{T}_{34}\int\limits_{-\infty}^{+\infty}d\tau_{34}
\dfrac{e^{-2i|\mathbf{q}|\tau_{34}}}{2|\mathbf{q}|}\int\dfrac{d^3\mathbf{k}}{(2\pi)^3}\mathbb{D}_{\mu\nu}\left[
f_{\mathbf{k}}(\tau_3)f^*_{\mathbf{k}}(\tau_4);
f^*_{\mathbf{k}-\mathbf{q}}(\tau_4)f_{\mathbf{k}-\mathbf{q}}(\tau_3)
\right].
\end{equation}
For $ |\mathcal{T}|<T $ we approximate the modes by (\ref{straight_modes}) as before and obtain:
\begin{equation}
\begin{split}
n_{\mu\nu}(\mathbf{q},\mathcal{T}) = -2e^2\int\limits_{-T}^{\mathcal{T}}d\mathcal{T}_{34}\int\limits_{-\infty}^{+\infty}d\tau_{34}
\dfrac{e^{-2i|\mathbf{q}|\tau_{34}}}{2|\mathbf{q}|}\int\dfrac{d^3\mathbf{k}}{(2\pi)^3}\times \\
\times \mathbb{D}_{\mu\nu}\left[
g_{\mathbf{k}_\perp}(k_1+eE\tau_3)g^*_{\mathbf{k}_\perp}(k_1+eE\tau_4);
g^*_{\mathbf{k}_\perp-\mathbf{q}_\perp}(k_1-q_1+eE\tau_4)g_{\mathbf{k}_\perp-\mathbf{q}_\perp}(k_1-q_1+eE\tau_3)
\right].
\end{split}
\end{equation}
Performing the shift  $ k_1\mapsto k_1+eE\mathcal{T}_{34} $ we finally get:
\begin{equation}
\begin{split}
n_{\mu\nu}(\mathbf{q},\mathcal{T}) = -2e^2(\mathcal{T}+T)\int\limits_{-\infty}^{+\infty}d\tau
\dfrac{e^{-2i|\mathbf{q}|\tau}}{2|\mathbf{q}|}\int\dfrac{d^3\mathbf{k}}{(2\pi)^3}\times \\
\times \widetilde{\mathbb{D}}_{\mu\nu}\left[
g_{\mathbf{k}_\perp}(k_1+eE\tau)g^*_{\mathbf{k}_\perp}(k_1-eE\tau);
g^*_{\mathbf{k}_\perp-\mathbf{q}_\perp}(k_1-q_1-eE\tau)g_{\mathbf{k}_\perp-\mathbf{q}_\perp}(k_1-q_1+eE\tau)
\right],
\end{split}
\end{equation}
which explicitly grows with $ \mathcal{T} $ and $T$. Hence, if we take into account the corrected propagator for photons in the two-loop correction to the current:
\\
\begin{figure}[h]
\vspace{-0ex}
\center{\includegraphics[width=0.4\linewidth]{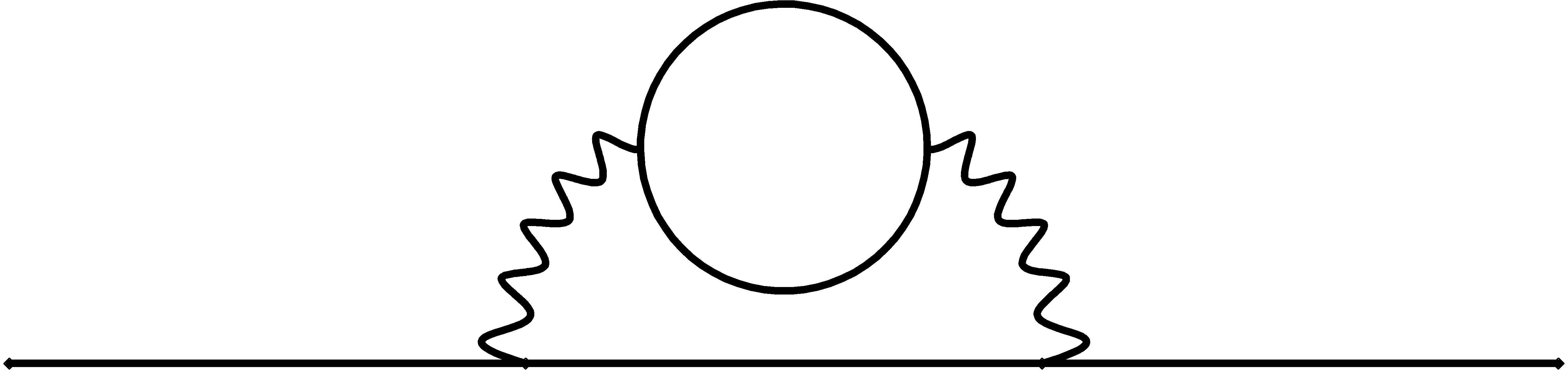}}
\vspace{-0ex}
\end{figure}
\\
we obtain the additional growth in $T$ for the two-loop corrected current.

\section{Discussion and Conclusion}
In this paper we have calculated one-loop correction to the scalar propagator and its contribution to the current of the created pairs in the lengthy and strong electric pulse. It turns out that the first loop correction in the scalar QED does not substantially alter the tree-level current. 

Furthermore, we also have calculated one-loop correction to the photon propagator and have shown that it gives an additional power of growth with the duration of the pulse for the current, $T$.

Despite the fact that one-loop diagrams do not bring an additional growth in $T$, analysis in the last section reveals that such a growth does have a place in higher loops. Furthermore, the following diagrams can also grow, for the same reason as why the loop correction to the photon propagator does grow:
\\
\begin{figure}[h]
\centering
\begin{subfigure}{.5\textwidth}
  \centering
  \includegraphics[width=0.7\linewidth]{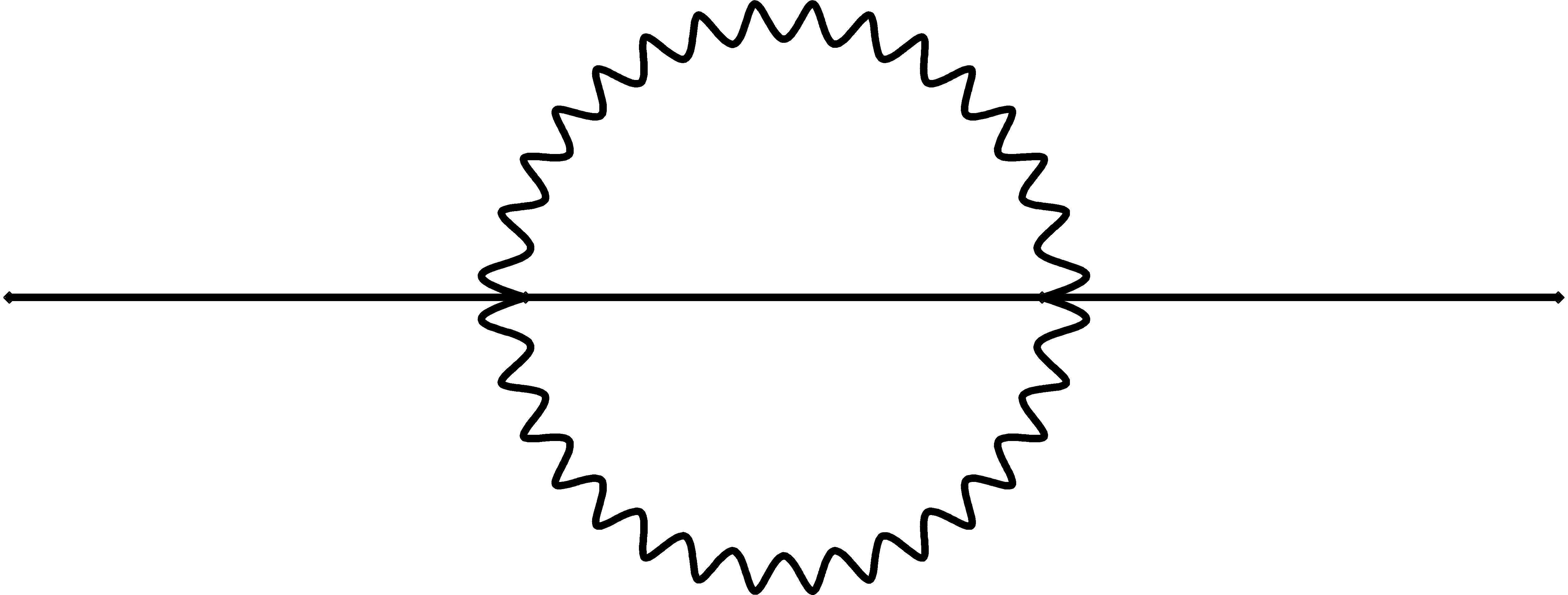}
  \caption{}
  \label{Sunsetphoton}
\end{subfigure}%
\begin{subfigure}{.5\textwidth}
  \centering
  \includegraphics[width=0.7\linewidth]{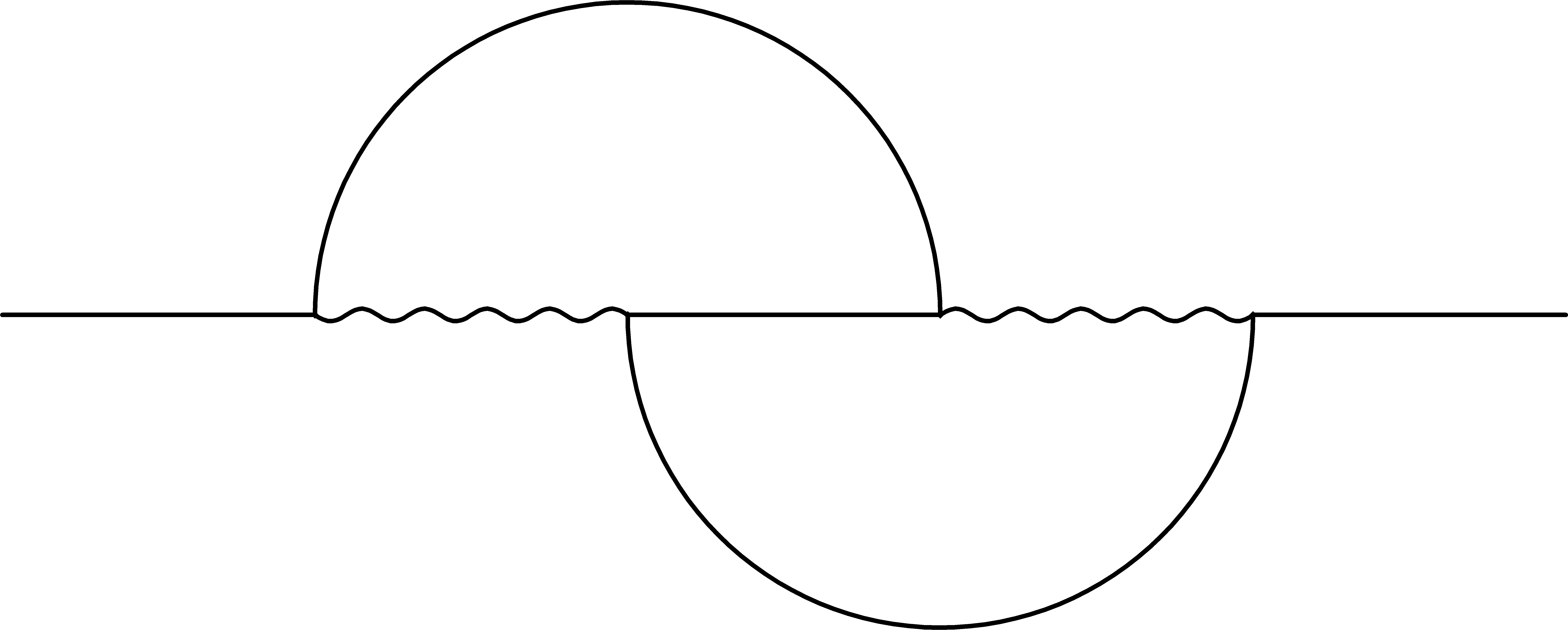}
  \caption{}
  \label{assympSunset}
\end{subfigure}
\caption{Two-loop diagrams with expected growth}
\label{fig:test}
\end{figure}
\\
The  diagram on the fig. \ref{Sunsetphoton} is expected to grow even with the tree-level photon propagator is used inside the loops, as the corresponding kinetic process is Thomson scattering. We believe that the diagram \ref{assympSunset} does grow with time. 

Another way to see the same effects is as follows. Due to the Debye screening\footnote{To observe such an effect we have to consider either pair creation in a medium or to wait long enough to create a gas of pairs.} the interaction of the scalars mediated by the photon exchange is localized, and can be approximated by the $\lambda\varphi^4$ self-interaction as low-energy effective theory. Then, if we shrink the photon lines in \ref{assympSunset} into points we will get simply sunset diagram of $\lambda\varphi^4$ theory, which does grow with time.

In all, we expect that loop corrections do change the tree-level Schwinger current of created pairs for a long enough pulse. The result of the change depends on initial conditions and has to be considered separately via a resummation of leading corrections from all loop levels.

\section{Acknowledgements}
We would like to acknowledge discussions with A. G. Semenov. This work was supported by RFBR grant 23-22-00145.

\section{Appendix}
\appendix
\section{Tadpole diagram}\label{Tadpole}
In this section we consider the tadpole diagram:
\\
\begin{figure}[h]
\center{\includegraphics[width=0.4\linewidth]{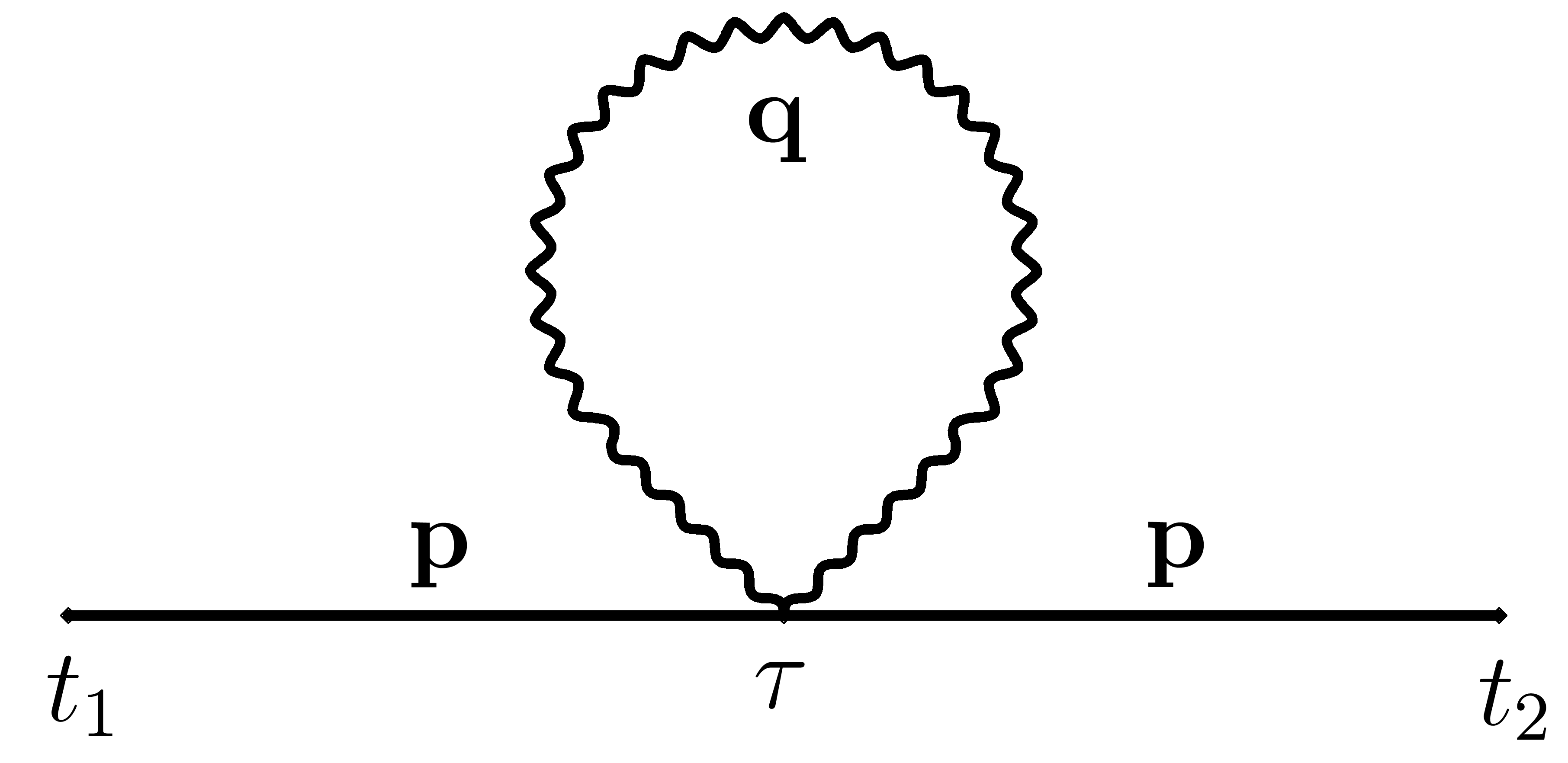}}
\caption{Tadpole diagram}
\label{image_tadpole_diagram}
\end{figure}
\\
It can be expected that this diagram brings an additional power of $T$ into the dependence of the current on the pulse duration $T$. However, as the photon propagator and the modes (\ref{a_decomposition}) are the same as in the situation without a background field, one directly obtains that
\begin{equation}\label{tadpole-propagator}
    \Delta_{\text{tad}}D^{ab}(\mathbf{p},t_1,t_2)=ie^2\mathcal{C}\sum\limits_{c}\sgn(c)\int d\tau D^{ac}(\mathbf{p},t_1,\tau)D^{cb}(\mathbf{p},\tau,t_2),
\end{equation}
so that the integration over the internal momentum $\mathbf{q} $ factorises to separate photon loop integral:
\begin{equation}\label{photon_loop_itself}
\mathcal{C}=\int\dfrac{d^3\mathbf{q}}{(2\pi)^3}\dfrac{1}{2|\mathbf{q}|}.
\end{equation}
The corresponding correction to the current has the following form:
\begin{equation}
j_1^{(\text{tad})}(t)=2e\int\dfrac{d^3\mathbf{p}}{(2\pi)^3}(p_1+eET)\left[
f^2_\mathbf{p}(t)\chi_\mathbf{p}(t)+\text{h.c.}\right];\quad
\chi_\mathbf{p}(t)=ie^2\mathcal{C}\int\limits_{t_0}^{t}\left[f^*_\mathbf{p}(\tau)\right]^2d\tau.
\end{equation}
Similarly to the other diagrams which have been considered earlier, the growing with the pulse duration contribution to $ j_1^{(\text{tad})}(t) $ can potentially come from the integration region over $ |\tau|<T $. Therefore, we can use the approximated modes (\ref{straight_modes}). Due to the proper Hadamard behavior of the exact scalar mode functions, we also restrict the integration over $ \mathbf{p} $ to $ |p_1|<eET $, considering $ p_1=0 $ in the leading order. In fact, high momentum modes do not feel the background field and, hence, behave as in empty space, i.e. just lead to the UV renormalization. As a result, we obtain
\begin{equation}\label{tadpole_current-1}
\begin{split}
j_1^{(\text{tad})} =& 4e^3\mathcal{C}\cdot eET\int\dfrac{d^2\mathbf{p}_\perp}{(2\pi)^3}eET
\\
\times & \left[
i\mathcal{A}_+(\mathbf{p})\mathcal{A}_-(\mathbf{p})\left(c^*(\mathbf{p}_\perp)\right)^2\int\limits_{-T}^{T}\left[
D^*_{-\frac{1}{2}+i\frac{\mathbf{p}^2_\perp+m^2}{2eE}}
\left[-\dfrac{(1-i)(p_1+eE\tau)}{\sqrt{eE}} \right]
\right]^2d\tau+\text{h.c.}
\right].
\end{split}
\end{equation}
Finally, we extend the integration in (\ref{tadpole_current-1}) over $\tau$ onto the entire real axis, which is valid up to the corrections of the order $ \mathcal{O}(1/T) $. Since $ eET\mathcal{A}_+(\mathbf{p})\mathcal{A}_-(\mathbf{p})\left(c^*(\mathbf{p}_\perp)\right)^2 $ does not depend on $ T $, one can see that $ j_1^{(\text{tad})} $ does not bring an additional positive power of $ T $ in comparison with (\ref{tree_current_renorm_finally}). Hence, the only effect is the UV renormalization of the bare constants in the theory by the loop integral $\mathcal{C}$, which can be removed by appropriate counterterms, so that the mass of the scalar in the bare Lagrangian gets renormalized.

\begin{figure}[h]
\center{\includegraphics[width=0.4\linewidth]{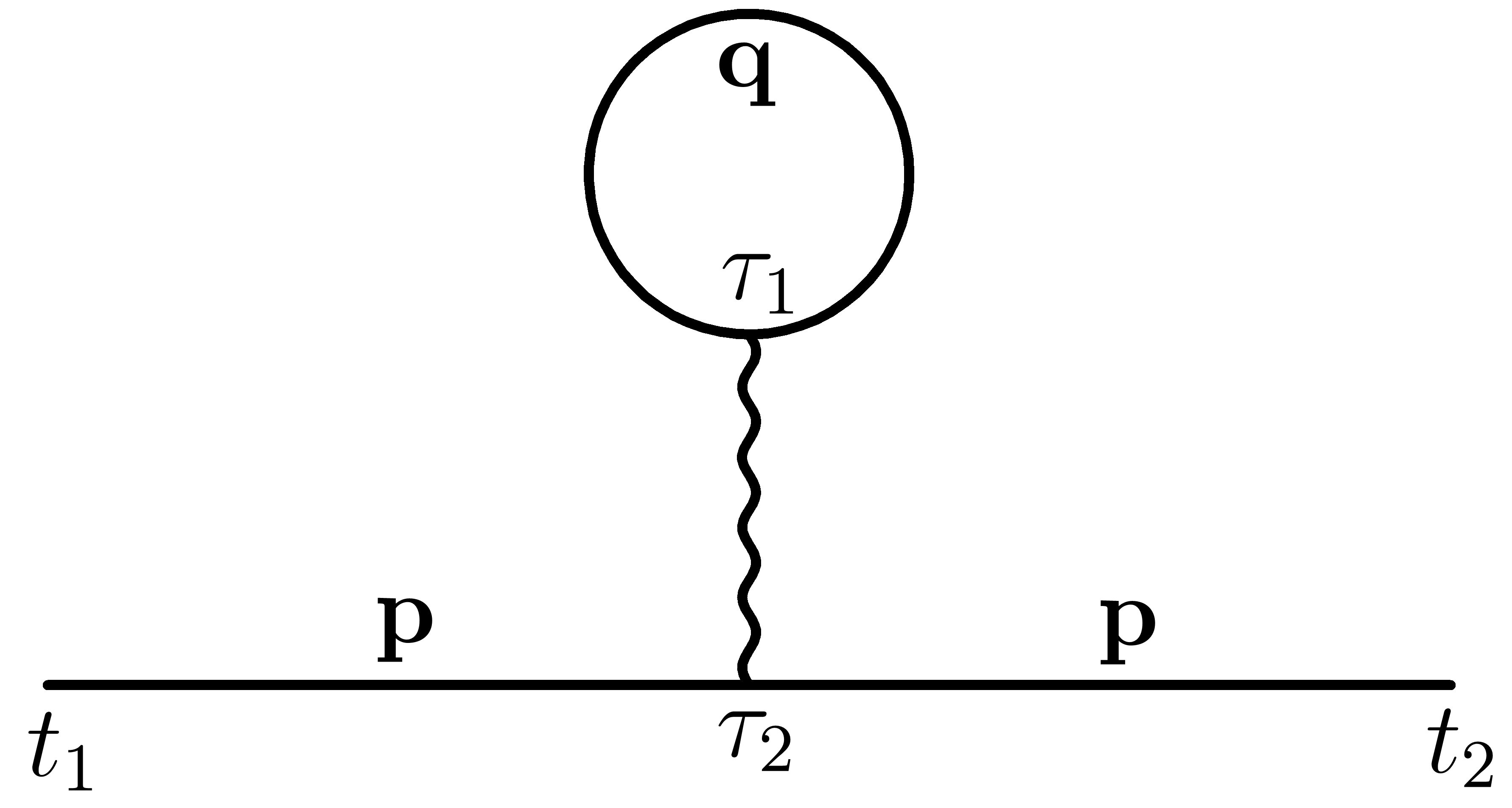}}
\caption{Another tadpole diagram}
\label{image_tadpole_diagram-2}
\end{figure}

There is another type of tadpole diagrams which is shown on the fig. \ref{image_tadpole_diagram-2}. Similar calculations that have been performed above for other diagrams show that this diagram does not lead to any additional growth of the current over $T$. This diagram also leads to a renormalization of the coupling constants in the theory.

\section{Time regions outside the pulse}\label{Time_regions_outside}
In this section we show that the integration outside the region $ |\tau_{1,2}|<T $ in (\ref{n-p(t)}) does not bring relevant contribution to (\ref{non_anomaly}).
\subsection{}\label{Time_regions_outside-n}
First of all, let us consider the region $ \tau_{1,2}>T$. It is useful to make the change of variables $\tau=\frac{\tau_1-\tau_2}{2},\mathcal{T}=\frac{\tau_1+\tau_2}{2}$, which run in the ranges $[\mathcal{T}-t; t-\mathcal{T}] $ and $[-\infty,t]$, correspondingly. As we are concerned about the large times $t=\frac{t_1+t_2}{2}$ we can extend the integration over $\tau$ onto the real axis due to the fast oscillations of the integrand in (\ref{n-p(t)}), so that
\begin{equation}\label{almost_n_p(t)}
n_\mathbf{p}(t)\simeq -4e^2\int\limits_{-\infty}^{t}d\mathcal{T}
\int\limits_{-\infty}^{+\infty}d\tau
\int\dfrac{d^3\mathbf{q}}{(2\pi)^3}
\dfrac{e^{2i|\mathbf{q}|\tau}}{2|\mathbf{q}|}\cdot
\mathbb{D}\bigg(
f^*_\mathbf{p}(\tau_1);\;
f^*_{\mathbf{p}-\mathbf{q}}(\tau_1)
f_{\mathbf{p}-\mathbf{q}}(\tau_2);\;
f_\mathbf{p}(\tau_2)
\bigg).
\end{equation}
Next, using the asymptotics (\ref{modes_asymp}) we can expand the product of the modes in (\ref{almost_n_p(t)}) in order to analyse oscillating terms:
\begin{equation}\label{16_exp}
\begin{split}
e^{2i|\mathbf{q}|\tau}
f^*_\mathbf{p}(\tau_1)f^*_{\mathbf{p}-\mathbf{q}}(\tau_1)
f_{\mathbf{p}-\mathbf{q}}(\tau_2)f_\mathbf{p}(\tau_2) \simeq &
\left|A_+B_+\right|^2 e^{2i(|\mathbf{q}|-\omega_+-\Omega_+)\tau} \\
+& \left|A_+B_-\right|^2 e^{2i(|\mathbf{q}|-\omega_++\Omega_+)\tau} \\
+& \left|A_-B_+\right|^2 e^{2i(|\mathbf{q}|+\omega_+-\Omega_+)\tau} \\
+& \left|A_-B_-\right|^2 e^{2i(|\mathbf{q}|+\omega_++\Omega_+)\tau} \\
+& A_+^*B_+^*A_-B_- e^{2i(-\omega_+-\Omega_+)\mathcal{T}}e^{2i|\mathbf{q}|\tau} \\
+& A_+^*B_-^*A_-B_+ e^{2i(-\omega_++\Omega_+)\mathcal{T}}e^{2i|\mathbf{q}|\tau} \\
+& A_-^*B_+^*A_+B_- e^{2i(+\omega_+-\Omega_+)\mathcal{T}}e^{2i|\mathbf{q}|\tau} \\
+& A_-^*B_-^*A_+B_+ e^{2i(+\omega_++\Omega_+)\mathcal{T}}e^{2i|\mathbf{q}|\tau} \\
+& |A_+|^2 B_+^*B_- e^{2i(|\mathbf{q}|-\omega_+)\tau}  e^{2i(-\Omega_+)\mathcal{T}} \\
+& |A_+|^2 B_-^*B_+ e^{2i(|\mathbf{q}|-\omega_+)\tau}  e^{2i(+\Omega_+)\mathcal{T}} \\
+& |A_-|^2 B_+^*B_- e^{2i(|\mathbf{q}|+\omega_+)\tau}  e^{2i(-\Omega_+)\mathcal{T}} \\
+& |A_-|^2 B_-^*B_+ e^{2i(|\mathbf{q}|+\omega_+)\tau}  e^{2i(+\Omega_+)\mathcal{T}} \\
+& |B_+|^2 A_+^*A_- e^{2i(|\mathbf{q}|-\Omega_+)\tau}  e^{2i(-\omega_+)\mathcal{T}} \\
+& |B_-|^2 A_+^*A_- e^{2i(|\mathbf{q}|+\Omega_+)\tau}  e^{2i(-\omega_+)\mathcal{T}} \\
+& |B_+|^2 A_-^*A_+ e^{2i(|\mathbf{q}|-\Omega_+)\tau}  e^{2i(+\omega_+)\mathcal{T}} \\
+& |B_-|^2 A_-^*A_+ e^{2i(|\mathbf{q}|+\Omega_+)\tau}  e^{2i(+\omega_+)\mathcal{T}},
\end{split}
\end{equation}
where we've denoted $A_\pm=\mathcal{A}_\pm(\mathbf{p}),\; B_\pm=\mathcal{A}_\pm(\mathbf{p-q}),\;
\omega_+=\omega_+(\mathbf{p}),\;
\Omega_+=\omega_+(\mathbf{p-q})$ and the action of the operator $\mathbb{D} $ results only in the additional factors before oscillating exponents. Now we note that the first eight terms in (\ref{16_exp}) after the integration over $\tau$ produce delta-functions with arguments $|\mathbf{q}|\pm\omega_+\pm\Omega_+$, which are never zero for $|\mathbf{q}|>0$, so that the integrals
\begin{equation}
\int \dfrac{d^3\mathbf{q}}{(2\pi)^3|\mathbf{q}|} \delta\left( |\mathbf{q}|\pm\omega_+\pm\Omega_+ \right)\ldots = 0.
\end{equation}
The other eight terms in (\ref{16_exp} after the integration over $\mathcal{T}$ contain exponents $ \exp\left(\pm 2i\Omega_+ t\right)$ or $ \exp\left(\pm 2i\omega_+ t\right)$ from the upper limit of the integration, $ \mathcal{T}=t $, and similar exponents from the lower limit. The contributions with exponents $\exp\left(\pm 2i\Omega_{+} t\right)$ are suppressed after the integration over $\mathbf{p}$ due to Riemann-Lebesgue lemma. Indeed, in the UV region in (\ref{non_anomaly}) after the integration over $\tau, \; \mathcal{T}$ and $\mathbf{q}$ the integrand can be estimated as
\begin{equation}
\sim \int\dfrac{d^3\mathbf{p}}{(2\pi)^3} p \cdot p^2\cdot\dfrac{1}{p^6} \cdot\dfrac{1}{\Omega_{+}}e^{\pm 2i\Omega_+ t}\sim \int\dfrac{d^3\mathbf{p}}{(2\pi)^3}\dfrac{1}{p^4}
e^{\pm 2i\Omega_+ t},
\end{equation}
where we used that the second-order operator $\mathbb{D}$ gives additional $p^2$, each coefficient in the approximation (\ref{modes_asymp}) behaves at least as $\frac{1}{p}$ in the UV regime and $\Omega_{+}\sim p$. Therefore, the integrand is the measurable function and the Riemann-Lebesgue lemma is applicable. The same we conclude about the oscillating terms produced by the exponent $\exp\left(\pm 2i\omega_+ t\right)$. The non-oscillating terms, which appear after the multiplication of this exponent by $|f_{\mathbf{p}}|^2$ in (\ref{non_anomaly}) cancel each other. In fact:
\[ 
|f_{\mathbf{p}}|^2\cdot|B_+|^2\delta\left(|\mathbf{q}|-\Omega_+\right)\left[
A_-^*A_+ \dfrac{e^{2i\omega_+t}}{2i\omega_+}+\text{h.c.}
\right] \sim |B_+A_+A_-|^2\delta\left(|\mathbf{q}|-\Omega_+\right)\left[
\dfrac{1}{2i\omega_+}+\text{h.c.} \right] = 0
\]
All the terms from the lower limit of the integration over $ \mathcal{T} $ also do not bring relevant contribution to (\ref{non_anomaly}). Apparently, similar considerations lead to the conclusion that there are no important terms from the past region $\mathcal{T}<-T$, where the mode function reduces to the single plane wave (\ref{asympt_past}). Hence there is no secular growth in one-loop correction to the current. Hence, only the integration over $ |\tau_{1,2}|<T $ in (\ref{n-p(t)}) can potentially bring a growing contribution to (\ref{non_anomaly}).

\subsection{}\label{Time_regions_outside-kappa}
Let us consider now $\kappa_\mathbf{p}(t)$ contribution from (\ref{kappa-p(t)}). Similarly to $n_\mathbf{p}(t)$ we make change of variables  $\tau=\dfrac{\tau_1-\tau_2}{2},\quad \mathcal{T}=\dfrac{\tau_1+\tau_2}{2}$, extend the integration region over $\tau$ onto the half of real axis and use that:
 \begin{equation}
\int\limits_{0}^{+\infty}e^{ixt}dt=\pi\delta(x)+i\cdot\mathrm{p.v.}\dfrac{1}{x}.
\end{equation}
Thus, we obtain:
\begin{equation}\label{kappa_begin}
\kappa_\mathbf{p}(t)\simeq 4e^2\int\limits_{t_0}^{t}d\mathcal{T}\int\limits_{0}^{+\infty}d\tau\int\dfrac{d^3\mathbf{q}}{(2\pi)^3}\dfrac{e^{i|\mathbf{q}|(\tau_1-\tau_2)}}{2|\mathbf{q}|}\cdot
\mathbb{D}\left(
f_\mathbf{p}(\tau_1);
f^*_{\mathbf{p}-\mathbf{q}}(\tau_1)
f_{\mathbf{p}-\mathbf{q}}(\tau_2);
f_\mathbf{p}(\tau_2)
\right)
\end{equation} 
In this appendix we want to consider the contribution from the time region $\mathcal{T}\in[T,t]$. For this region we use the asymptotics (\ref{modes_asymp}) and obtain an expression similar to (\ref{16_exp}). After the integration over $\tau$ we neglect terms with delta-functions, as it was done for $n_\mathbf{p}(t)$. Then we integrate over $\mathcal{T}$, neglect oscillating terms and obtain the following expression for the upper limit of integration:
\begin{equation}\label{anomaly_tau}
\begin{split}
(f^*_\mathbf{p}(t))^2\delta\kappa_\mathbf{p}(t)=-e^2\int\dfrac{d^3\mathbf{q}}{(2\pi)^3}\dfrac{1}{2|\mathbf{q}|}\times \\
\times\left[
\left[\mathbb{D}^j_j+\omega_+^2-\Omega_+^2\right]\cdot\dfrac{|A_+|^4\left|B_+\right|^2}{i(|\mathbf{q}|-\Omega_+)}
\dfrac{1}{i(+\omega_+)}\right. \\
+\left[\mathbb{D}^j_j+\omega_+^2-\Omega_+^2\right]\cdot\dfrac{|A_+|^4\left|B_-\right|^2}{i(|\mathbf{q}|+\Omega_+)}
\dfrac{1}{i(+\omega_+)} \\
+\left[\mathbb{D}^j_j+\omega_+^2-\Omega_+^2\right]\cdot\dfrac{|A_-|^4\left|B_+\right|^2}{i(|\mathbf{q}|-\Omega_+)}
\dfrac{1}{i(-\omega_+)} \\
+\left[\mathbb{D}^j_j+\omega_+^2-\Omega_+^2\right]\cdot\dfrac{|A_-|^4\left|B_-\right|^2}{i(|\mathbf{q}|+\Omega_+)}
\dfrac{1}{i(-\omega_+)} \\
+\left[\mathbb{D}^j_j-(\omega_++\Omega_+)^2\right]\cdot\dfrac{4|B_+|^2 |A_+A_-|^2}{i(|\mathbf{q}|-\Omega_++\omega_+)}\cdot
(t-T) \\
+\left[\mathbb{D}^j_j-(\omega_+-\Omega_+)^2\right]\cdot\dfrac{4|B_-|^2 |A_+A_-|^2}{i(|\mathbf{q}|+\Omega_++\omega_+)}\cdot
(t-T) \\
+\left[\mathbb{D}^j_j-(\omega_+-\Omega_+)^2\right]\cdot\dfrac{4|B_+|^2 |A_-A_+|^2}{i(|\mathbf{q}|-\Omega_+-\omega_+)}\cdot
(t-T) \\
+\left.\left[\mathbb{D}^j_j-(\omega_++\Omega_+)^2\right]\cdot
\dfrac{4|B_-|^2 |A_-A_+|^2}{i(|\mathbf{q}|+\Omega_+-\omega_+)}\cdot (t-T)
\right]
\end{split}
\end{equation}
We can see that the contribution to anomalous average, $ \delta\kappa_\mathbf{p}(t) $ from the region $[T,t]$, does grow with $ t $. However this growth is canceled if we substitute (\ref{anomaly_tau}) into (\ref{anomalous_current_initial}):
\begin{equation}
\begin{split}\label{anomaly-(t)}
\delta j^{\text{anom},t}_1(t)=-4e^3\int\dfrac{d^3\mathbf{p}}{(2\pi)^3}(p_1+eET)\int\dfrac{d^3\mathbf{q}}{(2\pi)^3}\dfrac{1}{4\omega^2_+|\mathbf{q}|}\times \\
\times\left[4(p_1+eET)^2-2q_1(p_1+eET)+4\mathbf{p}^2_\perp-2\mathbf{p}_\perp\mathbf{q}_\perp\right]\times \\
\times\left[
\dfrac{4|\mathbf{q}|\cdot|A_+B_+|^2}{|\mathbf{q}|^2-\Omega_+^2}+
\dfrac{|\mathbf{q}|\cdot|B_+|^2}{\omega_+\left(|\mathbf{q}|^2-\Omega_+^2\right)}+\dfrac{|A_+|^2}{\Omega_+(|\mathbf{q}|+\Omega_+)}+
\dfrac{1}{4\omega_+\Omega_+(|\mathbf{q}|+\Omega_+)}
\right].
\end{split}
\end{equation}
The last term in the big brackets is the same as in the current without the background field. It disappears after the integration because it is an odd function. The second term diverges as $ln |\mathbf{p}| $, and leads to the current renormalization. The first term gives contribution of the order $ O(1/T) $, while the third one --- of the order $ O(T) $. Hence, the contribution from the third term is the leading one in (\ref{anomaly-(t)}). However it is suppressed by $ e^2 $ as compared to (\ref{tree_current_renorm_finally}).

The contribution from the lower limit of the integration over $\mathcal{T}\in[T,t]$ in (\ref{kappa_begin}) is also $ O(T) $ and suppressed in comparison with the tree-level current (\ref{tree_current_renorm_finally}). Similar situation is the integration over $\mathcal{T}\in[t_0,-T]$. Hence, we conclude that the contribution to (\ref{anomalous_current_initial}) from the time region outside the pulse is suppressed in comparison with the tree-level current (\ref{tree_current_renorm_finally}).

\bibliography{bibliography}
\bibliographystyle{unsrt}
\end{document}